\documentclass[square]{ws-procs11x85}
\usepackage{multicol,float} 
\begin{document}

\title{PARTICLE ACCELERATION, MAGNETIC FIELD GENERATION, AND ASSOCIATED EMISSION IN COLLISIONLESS RELATIVISTIC JETS}

\author{K.-I. NISHIKAWA$^{*}$, Y. MIZUNO$^{\dagger}$, and G. J. FISHMAN$^{\dagger}$}

\address{Gamma-Ray Astrophysics, NSSTC, $^{\dagger}$NASA/MSFC\\
320 Sparkman Dr., Huntsville, AL 35805, USA \\  
$^{*}$E-mail: ken-ichi.nishikawa-1@nasa.gov\\
http://gammaray.nsstc.nasa.gov/$\sim$nishikawa}

\author{P. HARDEE}

\address{Department of Physics and  Astronomy, The University of Alabama, \\
Tuscaloosa, AL, 35487 USA} 

\begin{abstract}
Nonthermal radiation observed from astrophysical systems containing relativistic jets and shocks, e.g., active galactic nuclei (AGNs), gamma-ray bursts (GRBs), and Galactic microquasar systems usually have power-law emission spectra. Recent PIC simulations using injected relativistic electron-ion (electro-positron) jets show that acceleration occurs within the downstream jet. Shock acceleration is a ubiquitous phenomenon in astrophysical plasmas. Plasma waves and their associated instabilities (e.g., the Buneman instability, other two-streaming instability, and the Weibel instability) created in the shocks are responsible for particle (electron, positron,  and ion) acceleration. The simulation results show that the Weibel instability is responsible for generating and amplifying highly nonuniform, small-scale magnetic fields. These magnetic fields contribute to the electron's transverse deflection behind the jet head. The ``jitter'' radiation from deflected electrons has different properties than synchrotron radiation which assumes a uniform magnetic field. This jitter radiation may be important to understanding the complex time evolution and/or spectral structure in gamma-ray bursts, relativistic jets, and supernova remnants.\end{abstract}


\bodymatter

\begin{multicols}{2}
\section{Introduction}\label{aba:sec1}
The non-thermal spectrum of GRB sources is thought by most to arise in shocks which develop beyond the 
radius at which the relativistic fireball has become optically thin to $\gamma\gamma$ collisions\cite{piran99}. However, the observed spectra are hard, with a significant fraction of the energy above the 
$\gamma\gamma \rightarrow e^{+}e^{-}$ formation energy threshold.  A high compactness parameter can 
result in new pairs being formed outside the originally optically thin shocks responsible for the primary 
radiation\cite{thom00}. 
Radiation scattered by the external medium, as the collimated $\gamma$-ray front propagates through the ambient medium, would be decollimated, and, as long as $l \gtrsim 1$ 
absorbed by the primary beam. An $e^{\pm}$  pair cascade can then be produced as photons are 
back-scattered by the newly formed $e^{\pm}$  pairs and interact with other incoming seed photons\cite{thom00,belo02,belo05,mesz01,rami02,li03,kuma04}.
Here, $l=L\sigma_{\rm T}/4\pi r_{\rm l}m_{\rm e}c^{3}$, where $L$ is the total luminosity, and $r_{\rm r}$ is the characteristic source dimension. 
In this paper, we consider the plasma instabilities generated by rapid $e^{\pm}$  pair creation in GRBs. The injection of  $e^{\pm}$  pairs induces strong streaming motions in the ambient medium ahead of the forward shock. This sheared flow will be Weibel-like unstable\cite{medv99},
and if there is time before the shock hits, the resulting plasma instabilities will generate sub-equipartition quasi-static long-lived magnetic fields on the collisionless temporal and spatial scales across the $e^{\pm}$  pair-enriched region. This process is studied using three-dimensional kinetic simulations of monoenergetic and broadband pair plasma shells
interpenetrating an unmagnetized medium\cite{rami07}.

\begin{figure*}
\begin{center}
\psfig{file= 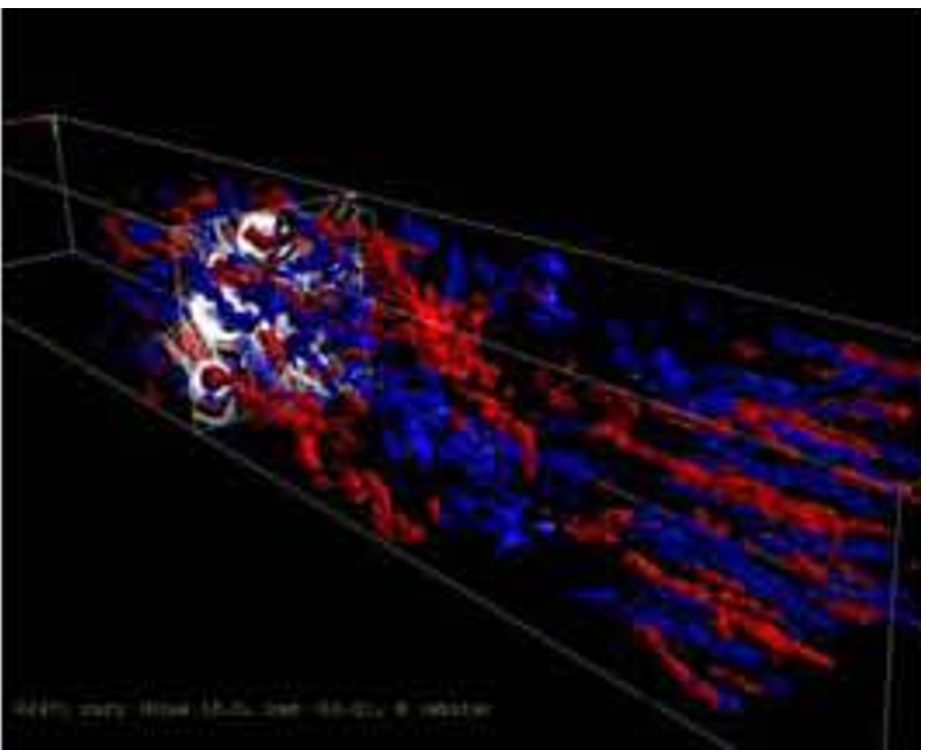,width=6cm}  \psfig{file= 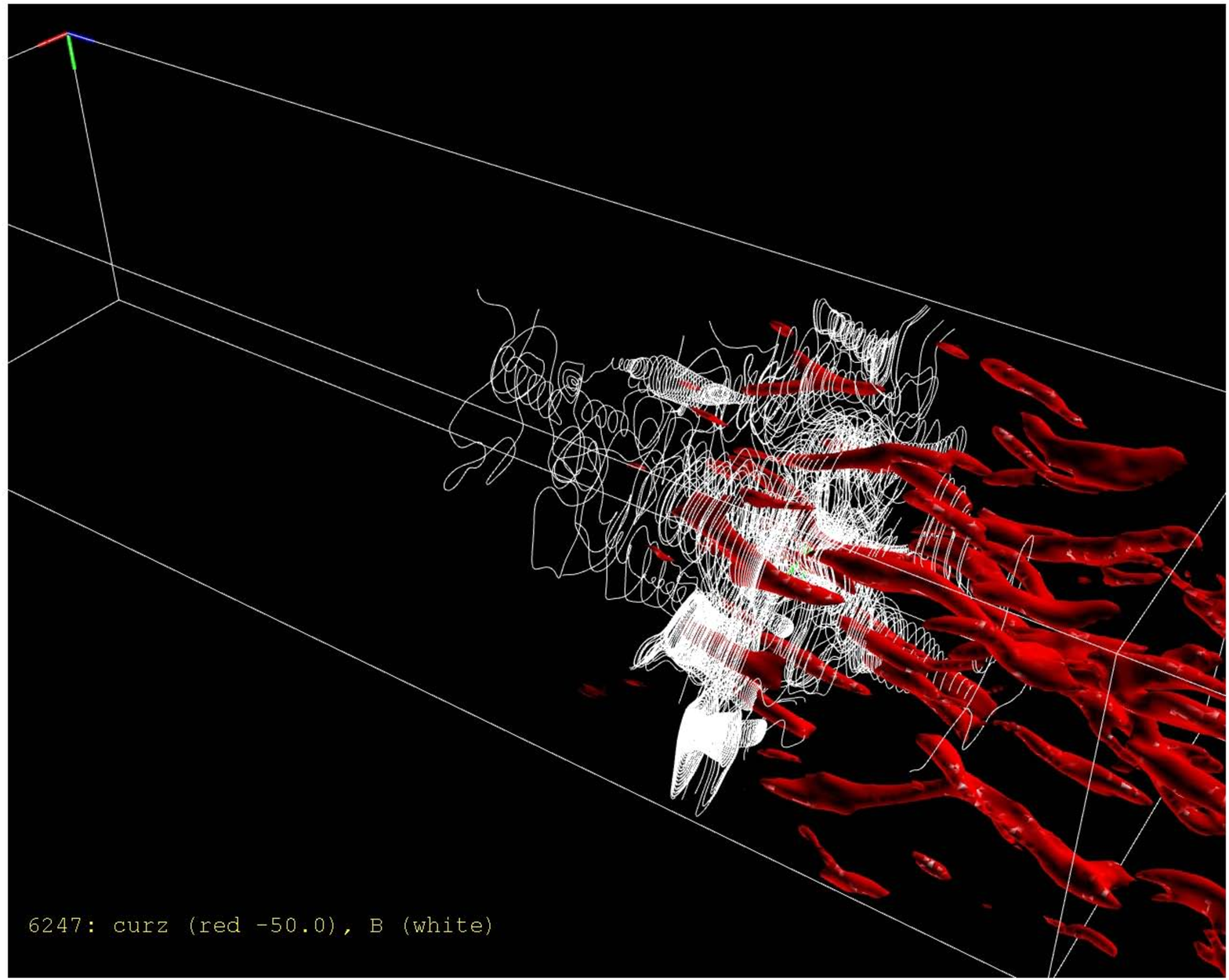,width=6cm}
\end{center}
\caption{Snapshots viewed from the
front of the jet at $t = 59.8/\omega_{\rm pe}$ left panel: the
isosurface of $Z$-component of the current density ($\pm J_{\rm z}$)
with the magnetic field lines (white) at the linear stage, and right
panel: the isosurface of $Z$-component of the current density
($-J_{\rm z}$) with the magnetic field lines behind the jet front.}
\end{figure*}

\section{Parameters and Initial Conditions used in Recent Simulations}

Simulations have been performed using an $85 \times 85
\times 640$ grid with a total of 380 million particles (27
particles$/$cell$/$species for the ambient plasma) and an electron
skin depth, $\lambda_{\rm ce} = c/\omega_{\rm pe} = 9.6\Delta$,
sufficient to study nonlinear spatial development\cite{bune93,nishi05,nishi06}. 
The time step $t = 0.013/\omega_{\rm pe}$ where $\omega_{\rm
pe} = (e^{2}n_{\rm e}/m_{\rm e})^{1/2}$ is the electron plasma
frequency ($n_{\rm e} = n_{\rm b}$ where $n_{\rm b}$ is the ambient
``background'' plasma density). 

\section{Monoenergetic \& Broad Pair Jet distributions
Injected into Pair \& Electron-Ion Plasmas}

The simulations described below were performed by a newly
parallelized OpenMP code on Columbia at NASA Advanced Supercomputing
(NAS). 

We have simulated four different initial pair jet distributions. Case
{\bf A} is a mono-energetic jet ($\gamma V_{\parallel} = 12.57c$)
injected into a pair ambient plasma and similar to published
simulations\cite{nishi05,nishi06}. 
Case {\bf B} is also a
mono-energetic jet but injected into an electron-ion ambient
plasma. Cases {\bf C} and {\bf D} correspond to pair jets with a broad
velocity distribution ($2 < u_{\parallel} = \gamma V_{\parallel}/c <
30$) created by photon annihilation\cite{mesz01}. In case {\bf C} the jet is injected into a pair ambient
plasma and in case {\bf D} the jet is injected into an electron-ion
ambient plasma.  For all cases the jet particles are very cold ($0.01
c$ in the rest frame). The average jet velocity for the broad pair
loaded distribution is $\gamma V_{\parallel}/c \sim 12.5$. The mass
ratio of electrons to ions in the ambient plasma is $m_{\rm i}/m_{\rm
e} = 20$. The electron thermal velocity in the ambient plasma is
$v_{\rm th} = 0.1c$, where $c$ is the speed of light. The ion thermal
velocity in the ambient plasma is $v_{\rm thi} = 0.022c$. Otherwise
parameters and setup are the same as those in the previous
simulations.

Encountering the medium at rest, incoming e$^{\pm}$ pairs are rapidly
deflected by field fluctuations.  The initial perturbations grow
nonlineary as the deflected e$^{\pm}$ pairs collect into current
channels. The resultant toroidal magnetic fields cause mutual
attraction between currents forcing like currents to approach each
other and merge. As a result, the magnetic field grows in
strength. This continues until the fields grow strong enough to
deflect the much heavier ions\cite{fred04}. The ions stay clearly separated in phase
space and are only slowly heated. In the presence of ions, the
magnetic field saturates at a higher level by a factor of $(m_{\rm
i}/m_{\rm e})^{1/2}=\sqrt{20}\sim 4.5$, albeit on a longer
timescale. In electron-ion jets, the real mass ratio of 1836 will be
expected to provide an even higher level of magnetic field and
enhanced electron acceleration\cite{hede05,heno05}. 

\begin{figure*}
\begin{center}
\psfig{file=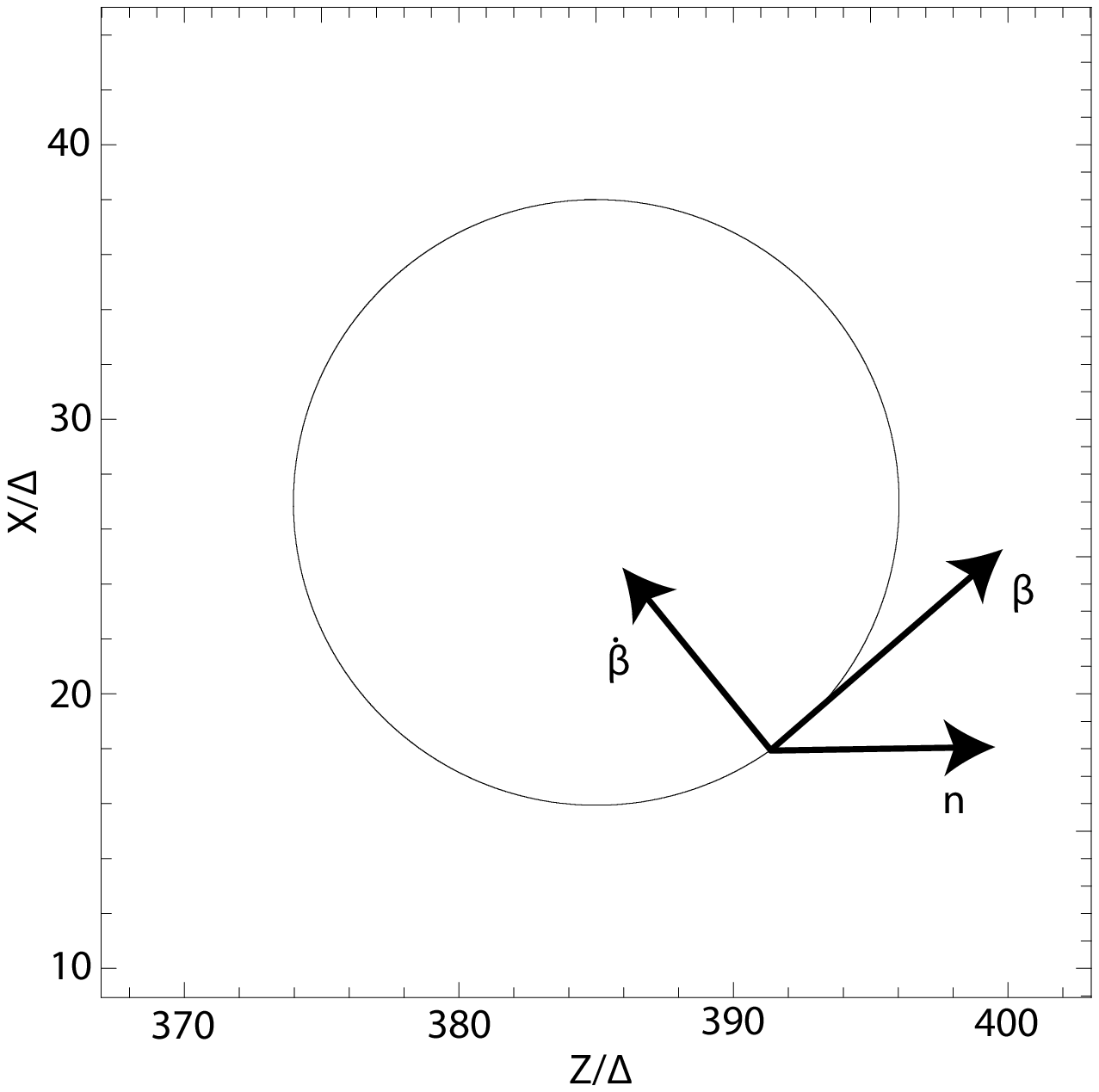,width=4.5cm}  \psfig{file=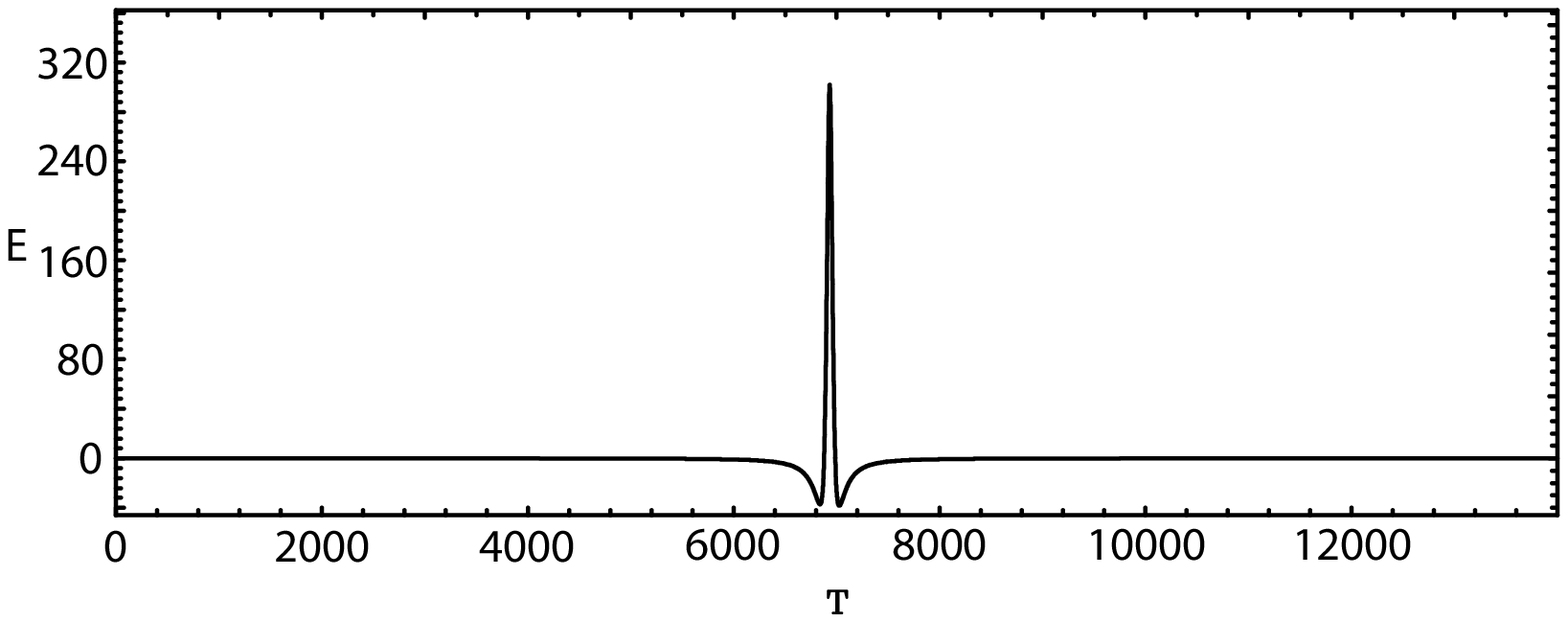,width=10.5cm}
\end{center}
\caption{The path of a charged particle moving in a homogenous magnetic
field (left panel). The particle radiates a time dependent electric field. An observer
situated at great distance along the n-vector sees the retarded electric field from
the gyrating particle (right panel). As a result of relativistic beaming, the field is
seen as pulses peaking when the particle moves directly towards the observer.}
\end{figure*}

The differences between an ambient consisting of pairs and an ambient
consisting of electrons and ions are due to the massive ion bulk
momentum constituting a vast energy reservoir for particle heating \cite{rami07}.
We also found that a broad
pair jet velocity distribution sustains a stronger magnetic field over
a longer region than a comparable mono-energetic pair distribution.
This additional effect is a result of the  more
``massive''  high Lorentz factor pairs in the jet.

The electron density and current filaments have a
complicated three-dimensional structure.  Current filaments ($J_{\rm
z}$) and associated magnetic fields (white curves) produced by the
filamentation (Weibel) instability form the dominant structures in the
relativistic collisionless shock shown in Figure 1.  In the linear
stage, the transverse size of these structures is nearly equal to the
electron skin depth but the longitudinal size (along the jet direction
as shown in the left side of the left panel) is much longer. The
growing current filaments in the linear stage merge in the
nonlinear stage (behind the jet front).  This appears spatially as
merging behind the jet front of the smaller scale filaments that
appear first.

\section{Radiation from a gyrating electron}
Let a particle be at position ${\bf{r}_{0}}(t)$ at
time $t$ (see Fig. 5 in Hededal \cite{hede05}).  At the same time, we observe the
electric field from the particle from position $\bf{r}$. However,
because of the finite propagation velocity of light, we observe the
particle at an earlier position $\bf{r}_{0}(t^{'})$ where it was at
the retarded time $t^{'} = t - \delta t^{'} = t - R(t^{'})/c$. Here
$R(t^{'}) = |\bf{r} - \bf{r}_{0}(t^{'})|$ is the distance from the
charge (at the retarded time $t^{'}$) to the observer point.

The retarded electric field from a charged particle moving with
instant velocity $\beta$ under acceleration $\dot{\beta}$ is
expressed as Jackson\cite{jack99}. 

\vspace*{-0.6cm}
\begin{eqnarray}
\bf{E} & = & \frac{q}{4\pi\epsilon_{0}}
\left[\frac{\bf{n}-\bf{\beta}}{\gamma^{2}
(1-\bf{\beta})^{\rm{3}}\rm{R}^{2}}\right]_{\rm ret} \nonumber\\
&&+
\frac{q}{4\pi\epsilon_{0}}
\left[\frac{\bf{n}\times{(\bf{n}-\bf{\beta})\times
\dot{\bf{\beta}}}}{(1-\bf{\beta})^{\rm{3}}\rm{R}}\right]_{\rm ret}
\end{eqnarray}

\vspace*{-0.3cm}
Here, $\bf{n} \equiv \bf{R}(t^{'})/ |R(t^{'})|$ is a unit vector
that points from the particles retarded position towards the
observer. The first term on the right hand side, containing the
velocity field, is the Coulomb field from a charge moving without
influence from external forces. The second term is a correction term
that arises, if the charge is subject to acceleration. Since the
velocity-dependent field is falling off as $R^{-2}$ while the
acceleration-dependent field falls off as $R^{-1}$, the latter
becomes dominant when observing the charge at large distances ($R
\gg 1$).

After some calculation and assumptions (for a detailed derivation see
e.g. Hededal\cite{hede05}) the total energy $W$ radiated per unit solid angle
per unit frequency is then expressed as

\vspace*{-0.6cm}
\begin{eqnarray}
&\frac{d^{2}W}{d\Omega d\omega} & =  \\
&\frac{\mu_{0} c
q^{2}}{16\pi^{3}} &\left|\int^{\infty}_{\infty}\frac{\bf{n}\times
[(\bf{n}-\mathbf{\beta})\times \dot{\bf{\beta}}]}{(1-\bf{\beta}\cdot
\bf{n})^{2}} 
 e^{i\omega(t^{'} -\bf{n} \cdot \bf{r}_{0}(t^{'})/c)}
dt^{'}\right|^{2} \nonumber
\end{eqnarray}
\vspace*{-0.5cm}

This equation contains the retarded electric field from a charged
particle moving with instant velocity $\beta$ under acceleration
$\dot{\beta}$, and only the acceleration field is kept since the
velocity field decreases by $1/R^{2}$. The distribution over frequencies
of the emitted radiation depends on the particle energy, radius of
curvature, and acceleration. These quantities are readily obtained
from the trajectory history of each charged particle.

\vspace*{-0.5cm}
\begin{figure}[H]
\center{\psfig{file=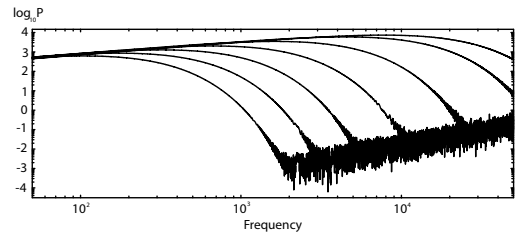,width=7cm}}
\caption{The observed power spectrum from a single charged particle, 
gyrating in a magnetic field with different viewing angles. The viewing angles are
0$^{\circ}$, 1$^{\circ}$, 
2$^{\circ}$, 3$^{\circ}$, 4$^{\circ}$, 5$^{\circ}$, and 6$^{\circ}$ ($n_{\rm y}\ne 0$) and their
peak frequencies are 7642,  4395,   1648,  666,    316,   166,   and  133, respectively.  With
larger angles the frequencies above the Nyquist frequency are strongly distorted.
The units on the axis are arbitrary.}
\end{figure}
\vspace*{-0.5cm}

The radiation from single electron gyrating in the uniform magnetic field ($B_{0} = B_{\rm y}$)
is calculated as shown in Fig. 2. The Lorenz factor of the electron is 40.8. The spectra observed far from the
electron in the $z$ direction are shown in Fig. 3. The critical frequency 
$f_{\rm c} = \frac{3}{2}\gamma^{3}\left(\frac{c}{\rho}\right) = 2309$, where $\rho = 11.03$.
The higher frequencies ($> f_{\rm c}$) are strongly damped with increasing
angles as   $e^{(-f/f_{\rm c})}$ in Jackson\cite{jack99}.

\section{Discussion}

We present self-consistent three-dimensional simulations of the fields developed by the electromagnetic filamentation instability as a consequence of e$^{\pm}$ pair injection, which arises naturally in 
the environment of GRB sources as a result of back-scattering.
Our results demonstrate that even in an initially unmagnetized scattering plasma, a small-scale, fluctuating, predominantly transversal, and near-equipartition magnetic field is unavoidably generated. 
These fields maintain a strong saturated level on spacial scales much longer than $\lambda_{\rm e}$ 
at least 
for the duration of the simulations $\sim 60(c/\lambda_{\rm e}) \sim 0.1(c/\lambda_{\rm T})$ . 
The e$^{\pm}$ pairs are effectively scattered with the magnetic field, thus effectively communicated their momentum to the scattering medium initially at rest. Our results indicate that the fields necessary to ensure 
that e$^{\pm}$ pairs remain coupled to the medium can be easily created via plasma
instabilities. The next required step is to increase the ion-to-electron mass ratio in the scattering medium 
in order to determine the spatial spread and character of the particle coupling

In order to obtain the spectrum of synchrotron (jitter) emission in future work 
\cite{medv00,medv06,work07,flei06},
we will consider an ensemble of selected electrons in the region where the
Weibel instability has grown fully and electrons are accelerated in
the generated magnetic fields. We will calculate emission from about
50,000 electrons during the sampling time, $t_{\rm s} = t_{\rm 2} -
t_{\rm 1}$ with Nyquist frequency $\omega_{\rm N} = 1/2\Delta t$ where
$\Delta t$ is the simulation time step and the frequency resolution
$\Delta \omega = 1/t_{\rm s}$. However, since the emission coordinate
frame for each particle is different, we accumulate radiation at a
fixed angle in simulation system coordinates after transforming from
the individual particle emission coordinate frame. This provides an
intensity spectrum as a function of angle relative to the simulation
frame $Z$-axis [can be any angle by changing the unit vector ${\bf n}$
in eq.\ (1)]. A hypothetical observer in the ambient medium (viewing
the external GRB shock) views emission along the system $Z$-axis. This
computation is carried out in the reference frame of the ambient
medium.  For an observer located outside the direction of bulk motion,
e.g., internal jet shocks or an external relativistic jet driven bow
shock, an additional Lorentz transformation is needed along the line
of sight to the observer. The spectra obtained from the simulations
need to be rescaled with a realistic time scale and relativistic
Doppler shift. In the electron-ion jets, the real mass ratio 1836 will
provide enhanced electron acceleration compared to a mass ratio of 20
\cite{hede05,heno05}.

\section*{Acknowledgments}

We have benefited from many useful discussions with E. Ramirez-Ruiz, J. Frederiksen, \AA.  Nordlund, 
and C. Hededal. This work is supported by  AST-0506719, AST-0506666, and NASA-NNG05GK73G. 
Simulations were performed on Columbia at NASA Advanced Supercomputing (NAS) and IBM p690 (Copper) 
at the National Center for Supercomputing Applications (NCSA) which is supported by the NSF. Part of 
this work was done while K.-I. N. was visiting the Niels Bohr Institute. He thank the director of the institution 
for its generous hospitality.

\end{multicols}
\end{document}